\documentclass[aps,prl,amsmath,amssymb,superscriptaddress,notitlepage,twocolumn]{revtex4-2}
\usepackage{amssymb, amsmath,amsbsy, array, latexsym, dsfont,layout, graphicx, mathrsfs, color,soul}
\usepackage{hyperref}

\newcommand{\ket}[1]{\left|{#1}\right\rangle}
\newcommand{\bra}[1]{\left\langle{#1}\right|}

\usepackage[normalem]{ulem}
\usepackage{bbold}
\usepackage{comment}

\begin{document}

\title{Universal, unambiguous concentration and distillation of Bell pairs}
\author{Orsolya K\'{a}lm\'{a}n}
\affiliation{HUN-REN Wigner Research Centre for Physics, 1525 P.O. Box 49, Hungary}
\author{Aur\'{e}l G\'{a}bris}
\affiliation{Faculty of Nuclear Sciences and Physical Engineering, Czech Technical University in Prague, B\v rehov\'a 7, 115 19 Praha 1 - Star\'e M\v esto, Czech Republic}
\affiliation{HUN-REN Wigner Research Centre for Physics, 1525 P.O. Box 49, Hungary}
\author{Igor Jex}
\affiliation{Faculty of Nuclear Sciences and Physical Engineering, Czech Technical University in Prague, B\v rehov\'a 7, 115 19 Praha 1 - Star\'e M\v esto, Czech Republic}
\author{Tam\'{a}s Kiss}
\affiliation{HUN-REN Wigner Research Centre for Physics, 1525 P.O. Box 49, Hungary}

\begin{abstract}
The ability of preparing perfect Bell pairs with a practical scheme is of great relevance for quantum communication as well as distributed quantum computing. We propose a scheme which probabilistically, but universally and unambiguously produces the $\ket{\Phi_{+}}$ Bell pair from four copies of qubit pairs initially in the same unknown pure quantum state. The same scheme, extended to eight qubit pairs initially in the same, moderately mixed quantum state, unambiguously produces the $\ket{\Phi_{+}}$ Bell pair with quadratically suppressed noise. The core step of the proposed scheme consists of a pair of local two-qubit operations applied at each of the two distant locations, followed by a partial projective measurement and postselection at each party, with results communicated classically. While the scheme resembles standard entanglement distillation protocols, it achieves success within just three iterations, making it attractive for real-world applications.
\end{abstract}

\maketitle

{\it Introduction ---} Bell pairs represent a quintessential form of entangled quantum states exhibiting nonlocality, regarded as a fundamental resource for quantum information processing and quantum communication. Preparation of Bell pairs and, more generally, entangled states in an LOCC (local operations and classical communication) setup has been an important problem in the past decades. In the literature, two different approaches are often distinguished for this task: entanglement concentration, and entanglement distillation or purification \cite{Dur_Briegel_2007}.

The goal of entanglement concentration is to prepare highly entangled states by appropriate local collective measurements on batches of $n$ pure states taken from an ensemble of identically prepared systems. The resulting maximally entangled multipartite state can then be transformed to standard Bell pairs with an efficiency approaching 1 for large $n$  \cite{Bennett_conc_1996,Hayashi_conc_2007}. This so-called Schmidt projection method is rather impractical for current physical implementations, as the two parties need to have large quantum memories both to store the qubits until they are transformed into Bell pairs and to use the produced Bell pairs for subsequent tasks such as teleportation. In a practical scheme, one could aim for a slightly different purpose: to minimize the size $n$ of the input batch, while requiring the probabilistic, but unambiguous preparation of a perfect Bell pair from an unknown, arbitrary input state. There are experiments demonstrating practical concentration of a certain set of unknown states \cite{Cao_2006,Xiong_2011}, however, to our knowledge, the problem with all the above requirements has not been explored yet. Although the Schmidt projection method can be executed for a lower number of pairs (two the lowest), by construction, it only guarantees unambiguous output if the Schmidt basis is known for the input.

Entanglement distillation addresses the task of preparing states with higher entanglement from mixed states \cite{Bennett_distill_1996,Bennett_distill_2_1996,Deutsch_distill_1996,Macchiavello_1998}. Usually, a maximally entangled Bell state is only achieved in the asymptotic limit with these methods. In real-world scenarios, however, relying on a large number of iterations may be prohibitive, as the number of input states scales exponentially with the number of iterative steps. Another general impractical feature of distillation protocols is that if they are applied to pure input states, they introduce noise (for example by the "twirling" step in the IBM protocol \cite{Bennett_distill_1996}, or by keeping two different measurement outcomes as in the Oxford protocol \cite{Deutsch_distill_1996}), which also disappears only asymptotically. Distillation procedures have since been further optimized \cite{Dur_Briegel_1998,Dur_Briegel_1999,Dur_Briegel_2003,Hsieh_2004,Torres_2016,Rozpedek_2018,Bernad_2022} and generalized e.g.\ to one-shot methods, aiming at producing approximate Bell pairs or other entangled states from a finite number of copies of the input state \cite{Brandao_2011,Buscemi_2010,Fang_2019}. In contrast to entanglement concentration \cite{Hayashi_conc_2007}, a generic caveat of distillation methods is that one must have certain a priori knowledge about the input state. 

Even though much is known about the mathematical limits of entanglement distillation under various assumptions, there remains a strong demand for practically applicable methods producing Bell pairs for quantum communication and distributed quantum computing \cite{Hickerson_2013, Monroe_2014}. Limitations from the point of view of practicality come from the lack of reliable quantum memories and the necessity to keep the processing transformations easily realizable.

Practically, it is plausible to process multiple input pairs in a single round in order to increase the number and quality of the produced Bell pairs. However, optimizing the performance of the overall distillation operation by numerical methods is getting hard already for 6-7 input pairs \cite{Krastanov}. Structured approaches employ few iterations of a well-designed, smaller LOCC operation for processing a higher number of input pairs in one round \cite{Kiss_2011,LOCCNet,Rozgonyi_2024}. 
A key challenge in analyzing these protocols lies in the inherently non-linear nature of the corresponding maps.
One class of protocols, e.g., the Extreme Photon Loss (EPL) method \cite{Bennett_EPL_1996} applies von Neumann measurements and accepts only one particular outcome in the post-selection, ensuring the advantageous property that pure input states are transformed into pure output states. In the EPL method a single CNOT operation is applied locally. Already if the CNOT is augmented by further single-qubit operations, the iterated dynamical behavior of the protocol becomes highly nontrivial \cite{Kiss_2011} even for single-qubit systems \cite{Kiss2006, Malachov2019}, due to the non-linear nature of the map. In general, slight changes in the unitary may lead to radically different behaviors \cite{Gilyen2016}. 


In this paper, we present LOCC protocols that can be used to concentrate entanglement from 4 pairs of qubits universally and unambiguously: no knowledge about the input state is required, and whenever the protocols succeed, the output is guaranteed to be the $\ket{\Phi_+}$ Bell state without the need for further postprocessing. The success probability is related to the concurrence of the initial state, consequently separable input states do not produce any output, while almost all entangled input states produce the designated output with some probability. We extend the study of these protocols to 8 slightly mixed initial input states, and show that they can unambiguously distill them into an almost perfect $\ket{\Phi_+}$ Bell pair, suppressing the initial noise quadratically. The protocols are based on the repeated applications of a core step, which consists a local two-qubit operation performed at each of the two parties. 

\begin{figure}
\begin{center}
\includegraphics[width=0.96\columnwidth]{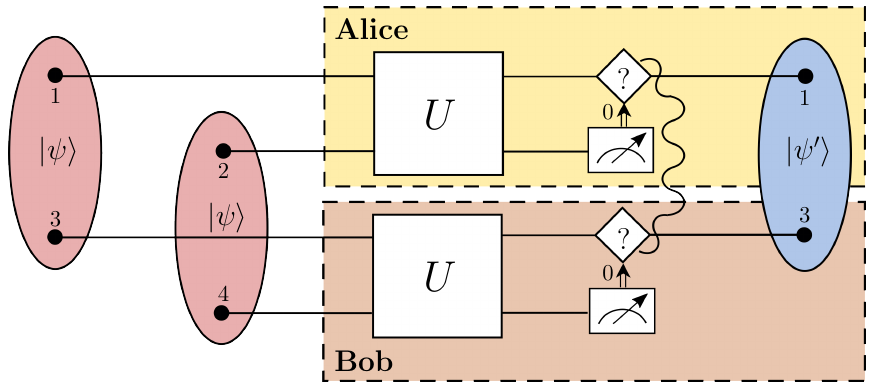} 
\vspace{-2ex}
\caption{The schematic representation of the core step of the protocol. $A$ and $B$ denote two distant parties, Alice and Bob, who apply a local two-qubit unitary $U$, and a subsequent measurement on qubits $2$ and $4$, respectively, after which they only keep qubits 1 and 3 if the measurements resulted 0, which they can communicate classically.}
\label{Fig0}
\end{center}
\end{figure}

\textit{The core step of the protocol ---} Let us assume that we are given an ensemble of qubit pairs in the pure quantum state $\ket{\psi}$, and take two pairs of qubits from this ensemble:
\begin{equation}
\ket{\psi}_{13}\!=\!\ket{\psi}_{24}\!=\!c_{1}\ket{00}+c_{2}\ket{01}+c_{3}\ket{10}+c_{4}\ket{11}, \label{psi0}
\end{equation}  
with  $\sum_{i=1}^{4}\left|c_{i}\right|^{2}=1$. Then, the first members of each pair (i.e., qubits $1$ and $2$) are sent to Alice, while the second members of the pairs (qubits $3$ and $4$) are sent to Bob (see Fig.~\ref{Fig0}). Alice and Bob both apply the same local two-qubit unitary operation $U$ to their qubits. After that, we assume that a measurement is performed on qubits $2$ and $4$, and qubits $1$ and $3$ are kept only if both measurements yielded $0$. For this latter decision to be made Alice and Bob use two-way classical communication. 

If both measurements succeed, then the state of qubits $1$ and $3$ is transformed into 
\begin{equation}
\ket{\psi'}_{13}=\mathcal{N}'\left[c\rq_{1}\ket{00}+c\rq_{2}\ket{01}+c\rq_{3}\ket{10}+c\rq_{4}\ket{11}\right]
\end{equation} 
where $\mathcal{N}'$ is a normalization factor which is related to the success probability of the protocol $P_{s}$ as $(\mathcal{N'})^{-2}=\sum_{i=1}^{4}\left|c_{i}'\right|^{2}=P_{s}$.

Let us assume that the unitary $U$ that is applied by Alice and Bob at both locations is one of the following unitaries:
\begin{equation}
U_{\pm}=\frac{1}{\sqrt{2}}\begin{bmatrix}
    0 & 1 & \pm 1 & 0 \\
    1 & 0 & 0 & \pm 1 \\
    0 & 1 & \mp 1 & 0 \\
    1 & 0 & 0 & \mp 1
\end{bmatrix},
\label{U1U2_00}
\end{equation}
which can be decomposed into elementary quantum gates as
\begin{align}
U_{+}&=(H\otimes\mathbb{1})U_{\mathrm{CNOT}}(\mathbb{1}\otimes X) \label{U+dec}\, , \\
U_{-}&=(X\otimes \mathbb{1})(H\otimes\mathbb{1})U_{\mathrm{CNOT}}(\mathbb{1}\otimes X)\, , \label{U-dec}
\end{align}
where $H$ is the Hadamard gate and $X$ is the Pauli-X gate. 

It can be shown that the transformation of the amplitudes (without normalization) after one step of the protocol is then given by
\begin{equation}
c_{1}\rq=c_{1}c_{4}\pm c_{2}c_{3}, \quad
c_{2}\rq=c_{3}\rq=0, \quad
c_{4}\rq=c_{1}c_{4}\mp c_{2}c_{3}. \label{1_step_trf}
\end{equation}
One can easily see that since the new amplitudes of the basis states $\ket{01}$ and $\ket{10}$ are zero irrespective of the initial amplitudes, quantum states which are of the form 
$\ket{\phi}=c_{1}\ket{00}+c_{4}\ket{11}$, $\left(|c_{1}|^{2}+|c_{4}|^{2}=1\right)$
are transformed into the maximally entangled Bell state $\ket{\Phi_{+}}=(\ket{00}+\ket{11})/\sqrt{2}$ already after one successful step if they contain any initial entanglement. This can be easily seen if one examines the success probability $P_{\mathrm{s}}^{(1)}=\left|c'_{1}\right|^{2}+\left|c'_{4}\right|^{2}=2\left|c_{1}c_{4}\right|^{2}$ of the step, which can be expressed using the concurrence, $C=2\left|c_{1}c_{4}-c_{2}c_{3}\right|=2\left|c_{1}c_{4}\right|$, as $P_{\mathrm{s}}^{(1)}=C^{2}/2$ in this case. Consequently, the less (more) entangled the initial state is, the smaller (higher) the success probability of the step is, thus the more (less) quantum systems we need to consume in order to have a successful single-step transformation.

Let us also note that quantum states of the form $\ket{\psi}=c_{2}\ket{01}+c_{3}\ket{10}$ become $\ket{\Phi_{-}}=(\ket{00}-\ket{11})/\sqrt{2}$ after one step, but if we apply the core step one more time, as we have mentioned above, $\ket{\Phi_{-}}$ becomes $\ket{\Phi_{+}}$.

\textit{Unambiguous concentration from 4 pairs of qubits} --- Surprisingly, the fact that a given pure initial state is transformed into $\ket{\Phi_{+}}$ is an even more general feature of the dynamics. This can be seen by calculating the output amplitudes after a second application of the core step on two copies of the already transformed $\ket{\psi'}$ state for a generic input, for which 
\begin{equation}
c_{1}''=c_{1}'c_{4}', \quad 
c_{2}''=c_{3}''=0, \quad
c_{4}''=c_{1}'c_{4}', \label{2_step_trf}
\end{equation}
independent of which of the unitaries $U_{\pm}$ was used in the protocol. Eq.~(\ref{2_step_trf}) describes the unnormalized $\ket{\Phi_{+}}$ Bell state as long as $c_{1}c_{4}\neq \pm c_{2}c_{3}$. Therefore, it is advantageous to consider the scheme involving two iterations of the core step where one inputs four copies of the initial two-qubit state $\ket{\psi}$: When all local measurements succeed, the remaining, unmeasured qubit pair will be unambiguously transformed into the $\ket{\Phi_{+}}$ Bell state.

In the following we show that for all states that do not yield the Bell state in Eq.~(\ref{2_step_trf}), the measurement will fail in the second step at the latest, making this protocol producing Bell states unambiguously. Input states that satisfy  $c_{1}c_{4}=c_{2}c_{3}$ are always separable, as their concurrence $C$ is zero. From Eq.~(\ref{1_step_trf}), one can see that after the first step either $c'_{4}=0$ and $c'_{1}=c_{1}c_{4}+c_{2}c_{3}$ (case of $U_{+}$), or $c'_{1}=0$ and $c'_{4}=c_{1}c_{4}+c_{2}c_{3}$ (case of $U_{-}$). Thus, whenever the first step succeeds, which happens with probability $P_{\mathrm s}^{(1)}=\left|c_{1}c_{4}+c_{2}c_{3}\right|^{2}=4\left|c_{1}c_{4}\right|^{2}$, then the input state is transformed into the separable state $\ket{00}$ or $\ket{11}$, respectively. If we apply a second step of the protocol on two copies of $\ket{00}$ or $\ket{11}$, then, as can be seen from Eqs.~(\ref{2_step_trf}), $c''_{1}=c''_{4}=0$, meaning that the second step never succeeds ($P_{\mathrm s}^{(2)}=\left|c''_{1}\right|^{2}+\left|c''_{4}\right|^{2}=0$). Thus, the two-step scheme does not produce an output for separable inputs.

Let us now consider inputs with $c_{1}c_{4}=-c_{2}c_{3}$, for which, as can be seen from Eqs.~(\ref{2_step_trf}), the two-step protocol also does not succeed. This condition represents a zero-measure set among the possible two-qubit input states, which we will call \textit{blind-spot inputs}, as these states 
can be entangled or even maximally entangled, yet, they "lie on the blind spot" of the protocol, as it cannot succeed when applied on them. The two different protocols (cases of $U_{+}$ and $U_{-}$) transform such states into $\ket{11}$ or $\ket{00}$, respectively, which are orthogonal to the case of product-state inputs. Thus, in principle, if one is given an ensemble of unknown states for which the second step of the protocol does not succeed for a large number of trials, one might differentiate the case of product-state inputs from blind-spot inputs by measuring the quantum states that were successfully transformed by the first step. If, for instance, in the $U_{+}$ case Alice and Bob measure their qubits to be $11$ after the first step (for sufficiently many trials) then they can assume that their initial ensemble was in a blind-spot state. Then, in order to  still be able to use such entangled states to produce $\ket{\Phi_{+}}$ Bell pairs, Alice and Bob can do the following: They can randomly choose a local single-qubit unitary each, and apply it on all of their qubits before the application of the protocol itself. These local operations do not change the entanglement, but they are likely to move the input states away from the $c_{1}c_{4}=-c_{2}c_{3}$ condition
so that the protocol can transform them into the $\ket{\Phi_{+}}$ Bell state if they initially contained some entanglement.   

In fact, the separable subset of the blind-spot states (for which both conditions are fulfilled, namely, $c_{1}c_{4}=\pm c_{2}c_{3}$), can be easily separated from the entangled blind-spot states. By examining these conditions it can be easily seen that separable blind-spot states are the ones, where at least one of the qubits is in a computational basis state. Interestingly, these inputs fail to produce an output already after the first step of the protocol, thus, they can be discriminated from entangled blind-spot inputs, which only fail at the second step.

\textit{The core step for mixed inputs} --- Let us now assume that the initial state of the ensemble is described by a general two-qubit mixed state $\rho$. We take two pairs of qubits with corresponding density matrices $\rho^{(1,3)}=\rho^{(2,4)}=\rho$, analogously to the pure case. Then, the initial four-qubit density matrix is given by $\rho^{(1,3)}\otimes\rho^{(2,4)}$. We consider the actions of $U_{\pm}$ of Eq.(\ref{U1U2_00}) at the locations of Alice (acting on qubits $1$ and $2$) and Bob (acting on qubits $3$ and $4$) by writing $\rho^{(1,3)}\otimes\rho^{(2,4)}$ in the computational basis where the qubits are ordered as ''$1,2,3,4$''
\begin{equation}
\tilde{\rho}=\left[U_{\pm}^{(1,2)}\otimes U_{\pm}^{(3,4)}\right]\left(\rho^{(1,3)}\otimes\rho^{(2,4)}\right)\left[U_{\pm}^{(1,2)}\otimes U_{\pm}^{(3,4)}\right]^{\dagger}.
\end{equation}
After the projection of qubits $2$ and $4$ onto the $\ket{0}$ state, the density matrix of qubits $1$ and $3$ transforms into a so-called $X$-state form. In the case of the protocol with $U_{+}$, the nonzero matrix elements of $\rho'$ (without renormalization) can be written as
\begin{align}
\rho'_{11}&=a_{+}+b_{+}+\mathrm{Re}\left(d_{+}\right), \notag \\
\rho'_{14}&=a_{-}+b_{-}- i\,\mathrm{Im}\left(d_{+}\right), \notag \\
\rho'_{22}&=a_{+}-b_{+}- \mathrm{Re}\left(d_{-}\right), \notag \\
\rho'_{23}&=a_{-}-b_{-}+ i\,\mathrm{Im}\left(d_{-}\right), \label{rhov} \\
\rho'_{33}&=a_{+}-b_{+}+ \mathrm{Re}\left(d_{-}\right), \notag \\
\rho'_{44}&=a_{+}+b_{+}- \mathrm{Re}\left(d_{+}\right), \notag 
\end{align}
with
\begin{align}
&a_{\pm}=\left(\rho_{11}\rho_{44}\pm\rho_{22}\rho_{33}\right)/2, \notag \\
&b_{\pm}=(\left|\rho_{14}\right|^{2}\pm\left|\rho_{23}\right|^{2})/2, \label{ai_bi} \\
&d_{\pm}=\rho_{12}\rho_{34}^{*}\pm\rho_{13}\rho_{24}^{*}. \notag
\end{align}
If one applies the protocol with $U_{-}$, then $\rho'$ is given by changing the sign of the last term in each matrix element in Eq.~(\ref{rhov}). 
The success probability of obtaining 0 as a measurement result for qubits $2$ and $4$, i.e., the success probability of transforming the initial state into the above $\rho'$ is given by ${\cal P}_{s}=\mathrm{Tr}(\rho')=4a_{+}$ in both cases.

\textit{Unambiguous distillation from 8 pairs of qubits} --- In what follows, we show that the protocol can unambiguously distill any moderately mixed state  into the $\ket{\Phi_{+}}$ Bell state with quadratically suppressed noise in three iterations of the core step. We express the initial state as
\begin{align}
\rho=(1-\varepsilon)\ket{\psi}\!\bra{\psi}+\varepsilon\rho_{\text{err}}=\ket{\psi}\!\bra{\psi}+\varepsilon M, \label{rho_arb_noise}
\end{align}
with $\ket{\psi}$ being an arbitrary (normalized) pure state (\ref{psi0}), $\rho_{\text{err}}$ is an arbitrary density operator describing the noise, $M=\rho_{\text{err}}-\ket{\psi}\!\bra{\psi}$ is a traceless Hermitian operator, and $\varepsilon\ll 1$. 

Without loss of generality, here we detail only the protocol involving the unitary $U_{+}$, however, the derivation for $U_{-}$ is then straightforward and yields essentially the same overall behavior. Using the transformation formulas (\ref{rhov}), one can determine the matrix elements of the unnormalized state $\rho'$ after one step of the protocol to be
\begin{align}
\rho'=\ket{\psi'}\bra{\psi'}+\varepsilon M'+\varepsilon^{2} Q ,
\label{rhov_pert}
\end{align}
where
$\ket{\psi'}$ is given by Eqs.~(\ref{1_step_trf}), and $M'$ and $Q$ can be easily determined from Eqs.(\ref{rhov}). The nonzero elements of $M'$ are
\begin{align}
&m'_{11}=\alpha_{+}+\text{Re}(\beta_{+})+\text{Re}(\gamma_{+}), \notag \\
&m'_{14}=\alpha_{-}+\text{Re}(\beta_{-})-i\text{Im}(\gamma_{+}), \notag \\
&m'_{22}=\alpha_{+}-\text{Re}(\beta_{+})-\text{Re}(\gamma_{-}), \notag \\
&m'_{23}=\alpha_{-}-\text{Re}(\beta_{-})+i\text{Im}(\gamma_{-}), \\
&m'_{33}=\alpha_{+}-\text{Re}(\beta_{+})+\text{Re}(\gamma_{-}), \notag \\
&m'_{44}=\alpha_{+}+\text{Re}(\beta_{+})-\text{Re}(\gamma_{+}), \notag
\end{align}
where
\begin{align}
&\alpha_{\pm}=\left(|c_{1}|^{2}m_{44}\pm|c_{2}|^{2}m_{33}\pm|c_{3}|^{2}m_{22}+|c_{4}|^{2}m_{11}\right)/2, \notag \\
&\beta_{\pm}=c_{1}c_{4}^{\ast}m_{14}^{\ast}\pm c_{2}c_{3}^{\ast}m_{23}^{\ast}, \label{alpha_beta_gamma} \\
&\gamma_{\pm}=c_{1}c_{2}^{\ast}m_{34}^{\ast}+c_{3}^{\ast}c_{4}m_{12}\pm c_{1}c_{3}^{\ast}m_{24}^{\ast}\pm c_{2}^{\ast}c_{4}m_{13}. \notag 
\end{align}
For simplicity, here we do not present the matrix $Q$, we only note that it has nonzero elements in its diagonal and anti-diagonal only, similarly to $M'$, and in accordance with Eq.~(\ref{rhov}). 

By taking $\rho'$ as the input of a second step, the resulting (unnormalized) state $\rho''$ can be written as
\begin{align}
\rho''=\ket{\Psi''}\bra{\Psi''}+\varepsilon M''+\mathcal{O}\left(\varepsilon^{2}\right) ,
\end{align}
where $\ket{\Psi''}=c'_{1}c'_{4}\ket{\Phi_{+}}$ is the unnormalized (pure) state after two iterations, and the nonzero matrix elements of $M''$ are
\begin{align}
&m''_{11}=m''_{14}=m''_{44}=\alpha'_{+}+\text{Re}(\beta'_{+}), \notag \\
&m''_{22}=m''_{23}=m''_{33}=\alpha'_{+}-\text{Re}(\beta'_{+}),
\end{align}
where
\begin{align}
&\alpha'_{+}=\left(|c'_{1}|^{2}m'_{44}+|c'_{4}|^{2}m'_{11}\right)/2,  \label{alphav_betav}\\
&\beta'_{+}=c'_{1}(c'_{4})^{\ast}(m'_{14})^{\ast}.
\end{align}

Now taking again $\rho''$ as the input of a third step, the resulting (unnormalized) state $\rho'''$ can be written as
\begin{align}
\rho'''=\ket{\psi'''}\bra{\psi'''}+\varepsilon M'''+\mathcal{O}\left(\varepsilon^{2}\right) ,
\end{align}
where $\ket{\psi'''}=\left(c'_{1}c'_{4}\right)^{2}\ket{\Phi_{+}}$ is the unnormalized (pure) state after three iterations, and the nonzero matrix elements of $M'''$ are
\begin{align}
&m'''_{11}=m'''_{14}=m'''_{44}=4|c'_{1}|^{2}|c'_{4}|^{2} m''_{11}, 
\end{align}
Note that $m'''_{22}, m'''_{23}, m'''_{33}$ have become zero in this step. Since $M'''$ is proportional to $\ket{\Phi_{+}}\bra{\Phi_{+}}$, and since $\varepsilon\ll 1$, it can be seen that after renormalization, $\rho'''$ is, in fact, to a good approximation, the pure state $\ket{\Phi_{+}}$. Thus, if the arbitrary pure input state is perturbed by any type of small noise, it will be purified into $\ket{\Phi_{+}}$ after three iterations of the core step, requiring 8 qubit pairs as inputs. 

Let us note that in the case of a perturbed Bell state, two iterations of the core step (4 pairs of inputs) is sufficient to produce a $\ket{\Phi_{+}}$ output with quadratically suppressed noise. This can be seen by substituting $c_{2}=c_{3}=0$ (or $c_{1}=c_{4}=0$) into Eqs.~(\ref{rhov_pert})-(\ref{alphav_betav}) from which, one can easily see that already after the second step  $M''\sim\ket{\Phi_{+}}\bra{\Phi_{+}}$. For product-state or blind-spot state inputs, 
even though there can be an output after the second step due to the contribution of the noise, the probability of an output after the third step will be quadratically suppressed. 

{\it Discussion ---} We have presented a practical distillation protocol based on a  low iteration-depth design. The structure of the protocol ensures that when the post-selection conditions are met, then the output can only be a high-fidelity $\ket{\Phi_+}$ Bell state. In contrast to usual entanglement distillation schemes, our protocol is agnostic to the input states, and no a priori information is required about the input state as long as the condition of small noise is met. Note that a rather detailed comparison with existing protocols is given in the Supplemental Material~\cite{SM}.

Our protocol has additional practically relevant features, as well. It enables the probabilistic preparation of $\ket{\Phi_+}$ Bell pairs using a small number of identically prepared entangled input states which may be pure or mixed containing a moderate amount of noise (the relative weight of the noise being significantly smaller than its pure state component). Depending on the quality of the input states, the number of required input states for a single shot of the protocol is at most 8 pairs, therefore it is not necessary to store the output of intermediate steps. The overall number of required operations (at most 7 two-qubit unitaries at one location) is relatively low, even in the case of noisy inputs, which is preferable from the point of view of minimizing decoherence and process noise. 

These properties make our protocol attractive for experimental realizations in quantum communication and distributed quantum computing. In particular, temporal delay loops may be used to implement the identical iteration steps in optical realizations protocols \cite{optical1, optical2, optical3, Ursin_2021, Peng_2021}. Additionally, our scheme could be readily applied to boost the efficiency of the recently proposed scheme of Jnane et al.\ for multi-core quantum computing \cite{Koczor_2022}, or the scaling up of quantum networks \cite{Kalb_2017}.

The above mentioned advantages combined with the ability to universally produce Bell pairs from arbitrary inputs make our scheme unique. While not directly comparable to other state-specific protocols in terms of efficiency, the flexibility and simplicity of the presented protocol render it experimentally feasible and a potentially useful tool in practice.

\section*{Acknowledgments}
We thank Saverio Pascazio, Stephen M. Barnett, J\'{a}nos Bergou, and G\'{a}bor Sz\'{e}chenyi  for many fruitful discussions. We also thank Maria Papanikolaou for the assistance in finding an earlier version of the gate decomposition. OK and TK acknowledge support from the National Research, Development and Innovation Office of Hungary, project No. TKP-2021-NVA-04.O.K. also acknowledges support from the European Union’s Horizon Europe research and innovation program under grant agreement No 101135699 (SPINUS). TK is grateful for the support from the ‘Frontline’ Research Excellence Programme of the NKFIH (Grant No. KKP133827). IJ and AG acknowledge the financial support from the European Union's research and innovation program under EPIQUE Project GA No 101135288, and from the state budget of the Czech Republic under RVO14000. IJ acknowledges support of the Grant Agency of the Czech republic (GA\v CR) under Grant No. 23-07169.

\end{document}